\documentclass[twocolumn,prl,showpacs]{revtex4}

\usepackage{dcolumn}
\usepackage{amsmath}
\usepackage{amssymb}

\begin{document}

\noindent \textbf{\large{Comment on ``Conservative Quantum Computing''}}

In \cite{Ozawa:02} Ozawa considers the limitations imposed by conservation
laws on the possibility of accurately generating quantum logic gates. In
particular, he argues that conservation of angular momentum imposes a
fundamental limit on how accurately the controlled-not (CNOT) gate can be
performed.
His argument runs roughly as follows. Since CNOT does not
commute with total angular momentum, it cannot be implemented directly in a
system in which angular momentum is conserved. Therefore one must enlarge
the two-qubit Hilbert space by attaching ancilla qubits, perform a unitary
operation on the resulting enlarged Hilbert space, then trace out the
ancillae. This defines a completely positive dynamical quantum map, which
can only yield the CNOT gate on the original two qubits \emph{probabilistically}. 
Ozawa calculates upper bounds on this probability of
success, and concludes that
the accuracy of quantum logic gates is fundamentally limited in the presence
of conservation laws.

The purpose of this comment is to point out that this is an overly restrictive conclusion. 
Indeed, as Ozawa notes at the end of his paper: ``The present investigation suggests that 
the current choice 
of the
computational basis should be modified so that the computational basis
commutes with the conserved quantity.'' ``... we may find such a
computational basis comprised of orthogonal entangled states over a
multiple-qubit system.'' Such a modification
of the computational basis is already well known. This is possible since while it is
true that the CNOT gate sometimes does not commute with the total angular momentum
operator, it does \emph{not} follow that CNOT cannot be generated from an
interaction that is rotationally invariant. It can, by using an encoding into multi-qubit 
states
that define an \emph{invariant subspace} with respect to the interaction at hand.
As first shown in \cite{Bacon:99a}, using an encoding of a logical qubit into four 
physical qubits, \emph{the
isotropic Heisenberg exchange interaction} $H_{\mathrm{Heis}}=J(X_{i}X_{j}+Y_{i}Y_{j}+
Z_{i}Z_{j})$, which commutes with total angular momentum,  
\emph{is universal for quantum computing}. Here $X_{i}$ is the Pauli $\sigma
_{x}$ matrix acting on qubit $i$, etc. This means that Heisenberg exchange can be
used to generate the group $U(2^K)$ on $K$ encoded qubits, and, in particular, it can
be used to generate an encoded CNOT gate, which treats only two out of the 16 dimensions 
of the 4-physical-qubit Hilbert space as computational basis states of an encoded qubit.
This result was soon followed by a
general proof that the Heisenberg exchange interaction is by itself
universal over encodings into any number of qubits $n\geq 3$ \cite{Kempe:00}. Specific gate sequences were then proposed for the $n=3$ qubit encoding 
\cite{DiVincenzo:00a}. Furthermore, by supplementing Heisenberg exchange
with a Zeeman splitting $\varepsilon (Z_{i}-Z_{j})$ (another interaction
that commutes with the total angular momentum operator), a simple encoding
into $n=2$ qubits, $|0_{L}\rangle =|0\rangle _{i}|1\rangle _{j}$ and $
|1_{L}\rangle =|1\rangle _{i}|0\rangle _{j}$, already suffices for universal
quantum computation \cite{Levy:01,WuLidar:01LidarWu:01}, and thus
to generate an encoded CNOT gate. In fact, other exchange interactions that
admit conserved quantities (e.g., the Hamiltonians $H_{\mathrm{XY}
}=J(X_{i}X_{j}+Y_{i}Y_{j})$ and $H_{\mathrm{XXZ}
}=J(X_{i}X_{j}+Y_{i}Y_{j})+J_{z}Z_{i}Z_{j}$, that have axial symmetry) are
also universal for quantum computation, with \cite{WuLidar:01LidarWu:01} or without 
Zeeman splitting \cite{Kempe:01WuLidar:JMP}. 

The reason that encoding helps is that it enables the quantum logic gates to
be executed on an invariant subspace of the full Hilbert space of the
original qubits plus ancillae. The exchange Hamiltonians, and hence the
encoded logic gates, preserve this subspace and, as is evident from their
universality, commute with all the symmetries of the system. Thus, Ozawa's
result can be seen as a confirmation of the advantage that may be had by allowing a flexible
definition of the computational basis: conservation laws do not impose fundamental
limitations on the accuracy of quantum logic operations, as long as one uses Hamiltonians
that act on an encoding of logical qubits into subspaces that are invariant under the 
symmetries
generating the conservation laws.

This work was sponsored by the DARPA-QuIST program
(managed by AFOSR under agreement No. F49620-01-1-0468). I am grateful to Prof. M. Ozawa for
helpful correspondence on some the issues raised here.

\noindent \begin{tabbing}
Da\=niel A. Lidar \\
  \>Chemical Physics Theory Group \\
  \>Chemistry Department\\
  \>80 St. George St.\\
  \>University of Toronto\\
  \>Toronto, Ontario, Canad M5S 3H6
\end{tabbing}

\end{document}